\documentclass[aps,showpacs,preprintnumbers,amsmath,amssymb]{revtex4}

\usepackage{float}
\usepackage{graphics,epsfig}
\usepackage{graphicx}
\usepackage{dcolumn}
\usepackage{bm}

\begin{document}
\baselineskip=0.8 cm
\title{{\bf Strong gravitational lensing in a squashed Kaluza-Klein black
hole spacetime}}
\author{Yue Liu}
\affiliation{Institute of Physics and Department of Physics,
Hunan Normal University,  Changsha, Hunan 410081, P. R. China \\
Key Laboratory of Low Dimensional Quantum Structures and Quantum
Control (Hunan Normal University), Ministry of Education, P. R.
China.}

\author{Songbai Chen}
\email{csb3752@163.com}  \affiliation{Institute of Physics and
Department of Physics,
Hunan Normal University,  Changsha, Hunan 410081, P. R. China \\
Key Laboratory of Low Dimensional Quantum Structures and Quantum
Control (Hunan Normal University), Ministry of Education, P. R.
China.}

\author{Jiliang Jing}
\email{jijing@hunnu.edu.cn} \affiliation{Institute of Physics and
Department of Physics,
Hunan Normal University,  Changsha, Hunan 410081, P. R. China \\
Key Laboratory of Low Dimensional Quantum Structures and Quantum
Control (Hunan Normal University), Ministry of Education, P. R.
China.}

\vspace*{0.2cm}
\begin{abstract}
\baselineskip=0.6 cm
\begin{center}
{\bf Abstract}
\end{center}

We investigate the strong gravitational lensing in a Kaluza-Klein
black hole with squashed horizons. We find the size of the extra
dimension imprints in the radius of the photon sphere, the
deflection angle, the angular position and magnification of the
relativistic images. Supposing that the gravitational field of the
supermassive central object of the Galaxy can be described by this
metric, we estimated the numerical values of the coefficients and
observables for gravitational lensing in the strong field limit.
\end{abstract}

 \maketitle
\newpage
\section{Introduction}

It is well known that string theory is a promising candidate for the
unified theory. The extra dimension is one of the most important
predictions in the string theory. Therefore, the detection of the
extra dimension has attracted  a lot of attention recently because
it can present the signature of the string and the correctness of
string theory.  Recent investigations indicate that the extra
dimension can imprint in the quasinormal modes originated from the
perturbation around high dimensional black holes \cite{Shen,
Abdalla,Chen,Kanti}, which could be tested in the gravitational wave
probe in the near future. Moreover, the spectrum of Hawking
radiation from the high dimensional black holes could provide
another possible way to observe the extra dimension, which is
expected to be detected in particle accelerator experiments \cite{5,
6, 7, 8, 9, 10, 11, 12, 13}.

The strong gravitational lensing is another possible way to detect
the extra dimension. According to general relativity, photons would
be deviated from their straight paths as they pass close to a
compact and massive body. The phenomena resulting from the
deflection of light rays in a gravitational field are called
gravitational lensing and the object causing a detectable deflection
is usually named a gravitational lens. The strong gravitational
lensing is caused by a compact object with a photon sphere, such as
a black hole, black brane, and so on. As photons pass close to the
photon sphere, the deflection angles become so large that an
observer would detect two infinite sets of faint relativistic images
on each side of the black hole which are produced by photons that
make complete loops around the black hole before reaching the
observer. These relativistic images can provide us not only
information about black holes in the Universe, but also profound
verification of alternative theories of gravity in their strong
field regime
\cite{Darwin,Vir,Vir1,Vir2,Vir3,Fritt,Bozza1,Eirc1,whisk,Bozza2,Bozza3,Bhad1,TSa1,AnAv}.
Thus, the study of the strong gravitational lensing by high
dimensional black holes can help us to detect the extra dimension in
astronomical observation in the future.

The Kaluza-Klein black holes with squashed horizons are a kind of
interesting Kaluza-Klein type metrics
\cite{IM,sq2,sq3,sq4,TVN,sq5,sq6}. In the vicinity of the black hole
horizon, the spacetime has a structure of the five-dimensional black
holes. In the region far from the black holes, it is locally the
direct product of the four-dimensional Minkowski spacetime and the
circle. Recently, the Hawking radiations have been studied in these
squashed Kaluza-Klein black holes, which indicates that the
luminosity of Hawking radiation can tell us the size of the extra
dimension which opens a window to observe extra dimensions
\cite{sq1hw,sq2hw}. Moreover, quasinormal modes in the background of
the Kaluza-Klein black hole with squashed horizons have been
investigated in \cite{sqq1,sqq2}, which implies that the quasinormal
frequencies contain the information of the size of the extra
dimension. Matsuno \textit{et al.}\cite{Ksq} have studied the
precession of a gyroscope in a circular orbit in the squashed
Kaluza-Klein black hole spacetime and find that the correction is
proportional to the square of (size of extra
dimension)/(gravitational radius of central object). These results
are very useful for us to understand the properties of the squashed
Kaluza-Klein black holes and to extract information about the extra
dimension in future observations. The main purpose of this paper is
to study the gravity lens in the strong field limit in the squashed
Kaluza-Klein black hole spacetime and see that effect of the size of
the extra dimension on the coefficients and observables of
gravitational lensing in the strong field limit.

The plan of our paper is organized as follows. In Sec.II we adopt to
Bozza's method \cite{Bozza2,Bozza3} and obtain the deflection angles
for light rays propagating in the squashed Kaluza-Klein black hole.
In Sec.III we suppose that the gravitational field of the
supermassive black hole at the center of our Galaxy can be described
by this metric and then obtain the numerical results for the
observational gravitational lensing parameters defined in Sec.II.
Then, we make a comparison between the properties of gravitational
lensing in the squashed Kaluza-Klein and four-dimensional
Schwarzschild metrics. To conclude, we present a summary.

\section{Deflection angle in the Squashed Kaluza-Klein Black
Hole spacetime}

The five-dimensional neutral static squashed Kaluza-Klein black hole
is described by \cite{IM,sq4}
\begin{eqnarray}
ds^2=-F(\rho)dt^2+\frac{K}{F(\rho)}d\rho^2+K\rho^2(d\theta^2+\sin^2\theta
d\phi^2)+\frac{r^2_{\infty}}{4K}(d\psi+\cos^2\theta d\phi)^2,
\label{metric0}
\end{eqnarray}
with
\begin{eqnarray}
F(\rho)=1-\frac{\rho_H}{\rho},\;\;\;\;\;\;\;\;\;\;K=1+\frac{\rho_0}{\rho},
\end{eqnarray}
where $0<\theta<\pi$, $0<\phi<2\pi$ and $0<\psi<4\pi$.  $\rho_H$ is
the radius of the black hole event horizon. The positive parameters
$r_{\infty}$, $\rho_0$ and $\rho_H$ are related by
$r^2_{\infty}=4\rho_0(\rho_H+\rho_0)$. The Komar mass of this black
hole is given by $M=\pi r_{\infty}\rho_H/G_5$ \cite{Ksq,Ksm}, where
$G_5$ is the five-dimensional gravitational constant. It is found
that in such a squashed Kaluza-Klein black hole spacetime the
relationship between $G_5$ and $G_4$ ( the four-dimensional
gravitational constant) can be expressed as $G_5=2\pi r_{\infty}G_4$
\cite{Ksq,Ksm}. Thus, the radius of the black hole event horizon
$\rho_H$ can be written as $\rho_H=2G_4M$ and the Hawking
temperature can be expressed as
\begin{eqnarray}
T_H=\frac{1}
{4\pi\rho_H}\sqrt{\frac{\rho_H}{\rho_H+\rho_0}}.\label{TH}
\end{eqnarray}
As $\rho_0\rightarrow 0$, the function $K=1$ and the metric
(\ref{metric0}) reduces to a spacetime which is described as the
four-dimensional Schwarzschild black hole with a constant twisted
$S_1$ fiber. As $\rho_H\ll\rho_0$, the function $K$ becomes
important, and the metric (\ref{metric0}) tends to a spherical
symmetrical five-dimensional Schwarzschild black hole
\cite{IM,sq4,Ksq}. Thus, the parameter $\rho_0$ can be regarded as a
scale of transition from five-dimensional spacetime to an effective
four-dimensional one.

The geodesics equations in the curved spacetime are
\begin{eqnarray}
\frac{d^2x^{\mu}}{d\lambda^2}+\Gamma^{\mu}_{\nu\tau}\frac{dx^{\nu}}{d\lambda}\frac{dx^{\tau}}{d\lambda}=0,
\end{eqnarray}
where $\lambda$ is an affine parameter along the geodesics. For the
squashed Kaluza-Klein black hole spacetime (\ref{metric0}), the
geodesics equations obey
\begin{eqnarray}
\baselineskip=1 cm
&&\frac{d^2t}{d\lambda^2}+\frac{F'(\rho)}{F(\rho)}\frac{dt}{d\lambda}\frac{d\rho}{d\lambda}=0,\label{ge1}\\
&&\frac{d^2\theta}{d\lambda^2}+\frac{(K\rho^2)'}{K\rho^2}\frac{d\theta}{d\lambda}\frac{d\rho}{d\lambda}-\cos\theta\sin\theta
\bigg(1-\frac{r^2_{\infty}}{4K^2\rho^2}\bigg)\bigg(\frac{d\phi}{d\lambda}\bigg)^2
+\frac{r^2_{\infty}\sin\theta}{4K^2\rho^2}\frac{d\phi}{d\lambda}\frac{d\psi}{d\lambda}=0,\label{ge2}\\
&&\frac{d^2\phi}{d\lambda^2}+\frac{(K\rho^2)'}{K\rho^2}\frac{d\phi}{d\lambda}\frac{d\rho}{d\lambda}+\frac{\cos\theta}{\sin\theta}
\bigg(2-\frac{r^2_{\infty}}{4K^2\rho^2}\bigg)\frac{d\phi}{d\lambda}\frac{d\theta}{d\lambda}
-\frac{r^2_{\infty}}{4K^2\rho^2\sin\theta}\frac{d\theta}{d\lambda}\frac{d\psi}{d\lambda}=0,\label{ge3}\\
&&\frac{d^2\psi}{d\lambda^2}+\frac{2(K\rho)'\cos\theta}{K\rho}\frac{d\phi}{d\lambda}\frac{d\rho}{d\lambda}
-\frac{K'}{K}\frac{d\psi}{d\lambda}\frac{d\rho}{d\lambda}
+\frac{1}{\sin\theta}
\bigg[1+\bigg(1-\frac{r^2_{\infty}}{4K^2\rho^2}\bigg)\cos^2\theta\bigg]\frac{d\phi}{d\lambda}\frac{d\theta}{d\lambda}
+\frac{r^2_{\infty}\cos\theta}{4K^2\rho^2\sin\theta}\frac{d\theta}{d\lambda}\frac{d\psi}{d\lambda}=0,\nonumber\\
\label{ge4}\\
&&\frac{d^2\rho}{d\lambda^2}+\frac{F(\rho)F(\rho)'}{2K}\bigg(\frac{dt}{d\lambda}\bigg)^2
+\frac{F(\rho)}{2K}\bigg(\frac{K}{F(\rho)}\bigg)'\bigg(\frac{d\rho}{d\lambda}\bigg)^2
-\frac{F(\rho)(K\rho^2)'}{2K}\bigg(\frac{d\theta}{d\lambda}\bigg)^2\nonumber\\
&&-\frac{F(\rho)}{2}\bigg(\frac{(K\rho^2)'\sin^2\theta}{K}\bigg)\bigg(\frac{d\phi}{d\lambda}\bigg)^2
+\frac{r^2_{\infty}K'}{8K^3}\bigg(\frac{d\phi}{d\lambda}\cos\theta+\frac{d\psi}{d\lambda}\bigg)^2
=0,\label{ge5}
\end{eqnarray}
where a prime represents a derivative with respect to $\rho$. In
this paper, we consider only the orbits with $\theta=\pi/2$. With
this condition, we find from (\ref{ge2}) that
\begin{eqnarray}
\frac{d\phi}{d\lambda}\frac{d\psi}{d\lambda}=0.
\end{eqnarray}
This implies that either $\frac{d\phi}{d\lambda}=0$ or
$\frac{d\psi}{d\lambda}=0$. Here we set $\frac{d\psi}{d\lambda}=0$,
so that we can compare with the results obtained in the
four-dimensional black hole spacetime. Obviously, Eq. (\ref{ge4}) is
satisfied naturally as $\frac{d\psi}{d\lambda}=0$. Therefore, the
equations of motion for this case can be simplified as
\begin{eqnarray}
&&\frac{dt}{d\lambda}=\frac{1}{F(\rho)},\nonumber\\
&&K\rho^2\frac{d\phi}{d\lambda}=J,\nonumber\\
&&\frac{d^2\rho}{d\lambda^2}+\frac{F(\rho)'}{2KF(\rho)}\bigg(\frac{dt}{d\lambda}\bigg)^2
+\frac{F(\rho)}{2K}\bigg(\frac{K}{F(\rho)}\bigg)'\bigg(\frac{d\rho}{d\lambda}\bigg)^2
-\frac{F(\rho)(K\rho^2)'J^2}{2K^3\rho^4}=0.
\end{eqnarray}
Following Refs.\cite{Darwin,Vir,Vir1,Vir2,Vir3,Fritt,Bozza1}, we can
obtain the deflection angle for the photon coming from infinite in
the squashed Kaluza-Klein black hole spacetime
\begin{eqnarray}
\alpha(\rho_{s})=I(\rho_{s})-\pi,
\end{eqnarray}
where $\rho_s$ is the closest approach distance and $I(\rho_s)$ is
\begin{eqnarray}
I(\rho_s)=2\int^{\infty}_{\rho_s}\frac{\sqrt{K}d\rho}{\sqrt{F(\rho)C(\rho)}
\sqrt{\frac{C(\rho)F(\rho_s)}{C(\rho_s)F(\rho)}-1}},\label{int1}
\end{eqnarray}
where $C(\rho)=K\rho^2=\rho(\rho+\rho_0)$. As in the
four-dimensional black hole spacetime, the deflection angle
increases when parameter $\rho_s$ decreases. When the deflection
angle becomes $2\pi$, the light ray make a complete loop around the
compact object before reaching the observer. If $\rho_s$ is equal to
the radius of the photon sphere, one can find that the deflection
angle diverges and the photon is captured. In the squashed
Kaluza-Klein black hole spacetime,  one can find from
Eq.(\ref{int1}) that the deflection angle contains the information
about the scale of transition $\rho_0$. This implies that we could
detect the extra dimension by the gravitational lens.

For the squashed Kaluza-Klein black hole spacetime, the photon
sphere equation is given by
\begin{eqnarray}
\frac{C(\rho)'}{C(\rho)}=\frac{F(\rho)'}{F(\rho)}.\label{root}
\end{eqnarray}
The radius of the photon sphere is the largest real root of Eq.
(\ref{root}), which can be expressed as
\begin{eqnarray}
\rho_{hs}=\frac{3\rho_H-\rho_0+\sqrt{\rho_0^2+10\rho_0\rho_H+9\rho^2_H}}{4}.\label{phs1}
\end{eqnarray}
Obviously, it also depends on the scale of transition $\rho_0$. As
$\rho_0$ approaches zero, the radius of the photon sphere
$\rho_{hs}=\frac{3}{2}\rho_H$, which is consistent with that in the
four-dimensional Schwarzschild black hole spacetime. As $\rho_0$
tends to infinite, we find that $\rho_{hs}\rightarrow2\rho_H$.
According to the coordinate translation
$\rho=\rho_0\frac{r^2}{r^2_{\infty}-r^2}$, one can obtain easily
that $r_{hs}\rightarrow\sqrt{2}r_H$ in the case that $\rho_0$ tends
to infinite,  which agrees with the photon sphere in the
five-dimensional Schwarzschild black hole spacetime. In Fig.(1), we
plot the variety of the ratio between the radius of the photon
sphere $\rho_{hs}$ and the radius of the black hole event horizon
$\rho_H$ with the parameter $\rho_0/\rho_H$. From fig.(1), one can
obtain that $\rho_{hs}/\rho_H$ increases with $\rho_0/\rho_H$.
\begin{figure}[ht]
\begin{center}
\includegraphics[width=7cm]{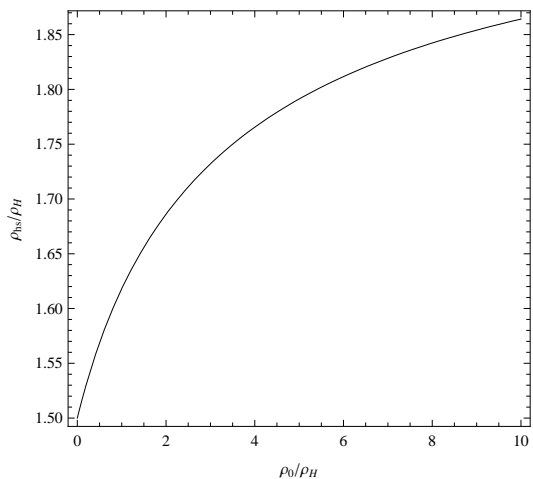}
\caption{Variety of the quantity $\rho_{hs}/\rho_H$ with
$\rho_0/\rho_H$ in the squashed Kaluza-Klein black hole spacetime.}
\end{center}
\end{figure}

Following the method developed by Bozza \cite{Bozza2}, we can define
a variable
\begin{eqnarray}
z=1-\frac{\rho_s}{\rho},
\end{eqnarray}
and rewrite  Eq.(\ref{int1}) as
\begin{eqnarray}
I(\rho_s)=\int^{1}_{0}R(z,\rho_s)f(z,\rho_s)dz,\label{in1}
\end{eqnarray}
with
\begin{eqnarray}
R(z,\rho_s)&=&2\frac{\rho^2\sqrt{K
C(\rho_s)}}{\rho_sC(\rho)}=2\sqrt{\frac{\rho_s+\rho_0}{\rho_s+\rho_0(1-z)}},
\end{eqnarray}
\begin{eqnarray}
f(z,\rho_s)&=&\frac{1}{\sqrt{F(\rho_s)-F(\rho)C(\rho_s)/C(\rho)}}.
\end{eqnarray}
Obviously, the function $R(z, \rho_s)$ is regular for all values of
$z$ and $\rho_s$. While the function $f(z, \rho_s)$ diverges as $z$
tends to zero. Thus, we split the integral (\ref{in1}) into the
divergent part $I_D(\rho_s)$ and the regular one $I_R(\rho_s)$
\begin{eqnarray}
I_D(\rho_s)&=&\int^{1}_{0}R(0,\rho_{hs})f_0(z,\rho_s)dz, \nonumber\\
I_R(\rho_s)&=&\int^{1}_{0}[R(z,\rho_s)f(z,\rho_s)-R(0,\rho_{hs})f_0(z,\rho_s)]dz
\label{intbr}.
\end{eqnarray}
\begin{figure}[ht]
\begin{center}
\includegraphics[width=7cm]{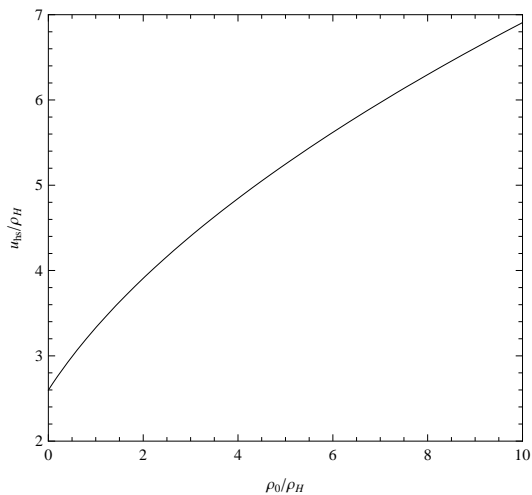}
\caption{The ratio $u_{hs}/\rho_H$ (between the minimum impact
parameter $u_{hs}$ and $\rho_H$)  changes with $\rho_0/\rho_H$ in
the squashed Kaluza-Klein black hole spacetime. }
\end{center}
\end{figure}
In order to find the order of divergence of the integrand, we expand
the argument of the square root in $f(z,\rho_{s})$ to the second
order in $z$
\begin{eqnarray}
f_s(z,\rho_{s})=\frac{1}{\sqrt{p(\rho_{s})z+q(\rho_{s})z^2}},
\end{eqnarray}
where
\begin{eqnarray}
p(\rho_{s})&=&2-\frac{2\rho_H}{\rho_{s}}-\frac{\rho_H+\rho_0}{\rho_{s}+\rho_0},  \nonumber\\
q(\rho_{s})&=&\frac{(\rho^2_0+3\rho_0\rho_s)\rho_H+(3\rho_H-\rho_s)\rho^2_s}{\rho_s
(\rho_0+\rho_s)^2}.
\end{eqnarray}
If $\rho_{s}$ tends to the radius of the photon sphere $\rho_{hs}$,
one can find that the coefficient $p(\rho_{s})$ approaches zero and
the leading term of the divergence in $f_s(z,\rho_{s})$ is $z^{-1}$,
which means that the integral (\ref{in1}) diverges logarithmically.
Close to the divergence, the deflection angle can be expanded in the
form \cite{Bozza2}
\begin{eqnarray}
\alpha(\theta)=-\bar{a}\log{\bigg(\frac{\theta
D_{OL}}{u_{hs}}-1\bigg)}+\bar{b}+O(u-u_{hs}), \label{alf1}
\end{eqnarray}
with
\begin{eqnarray}
&\bar{a}&=\frac{R(0,\rho_{hs})}{\sqrt{q(\rho_{hs})}}=\bigg[\frac{1}{2}\bigg(1+3\sqrt{\frac{\rho_0+\rho_H}{\rho_0+9\rho_H}}\bigg)\bigg]^{1/2}, \nonumber\\
&\bar{b}&=
-\pi+b_R+\bar{a}\log{\frac{\rho^2_{hs}[C''(\rho_{hs})F(\rho_{hs})-C(\rho_{hs})F''(\rho_{hs})]}{u_{hs}
\sqrt{F^3(\rho_{hs})C(\rho_{hs})}}}
=-\pi+b_R+\log{\bigg[2\sqrt{\frac{\rho_0+9\rho_H}{\rho_0+ \rho_H}}\;\bigg]}, \nonumber\\
&b_R&=I_R(\rho_{hs}), \nonumber\\
&u_{hs}&=\sqrt{\frac{C(\rho_{hs})}{F(\rho_{hs})}}=\frac{\sqrt{2}}{4}\bigg[27\rho^2_H+18\rho_0\rho_H-\rho^2_0+\sqrt{\rho_0+\rho_H}(\rho_0+9\rho_H)^{3/2}\bigg]^{1/2}.\label{coa1}
\end{eqnarray}
\begin{figure}[ht]
\begin{center}
\includegraphics[width=7cm]{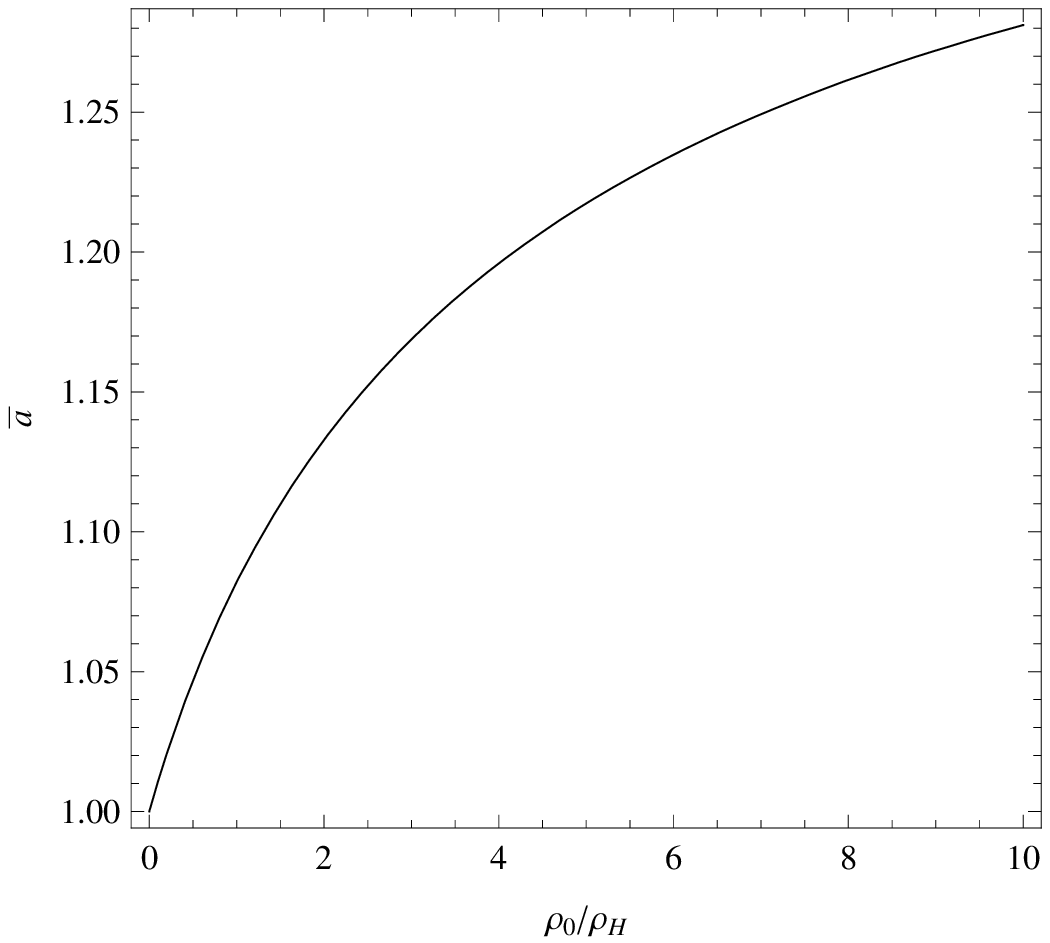}\;\;\;\includegraphics[width=7cm]{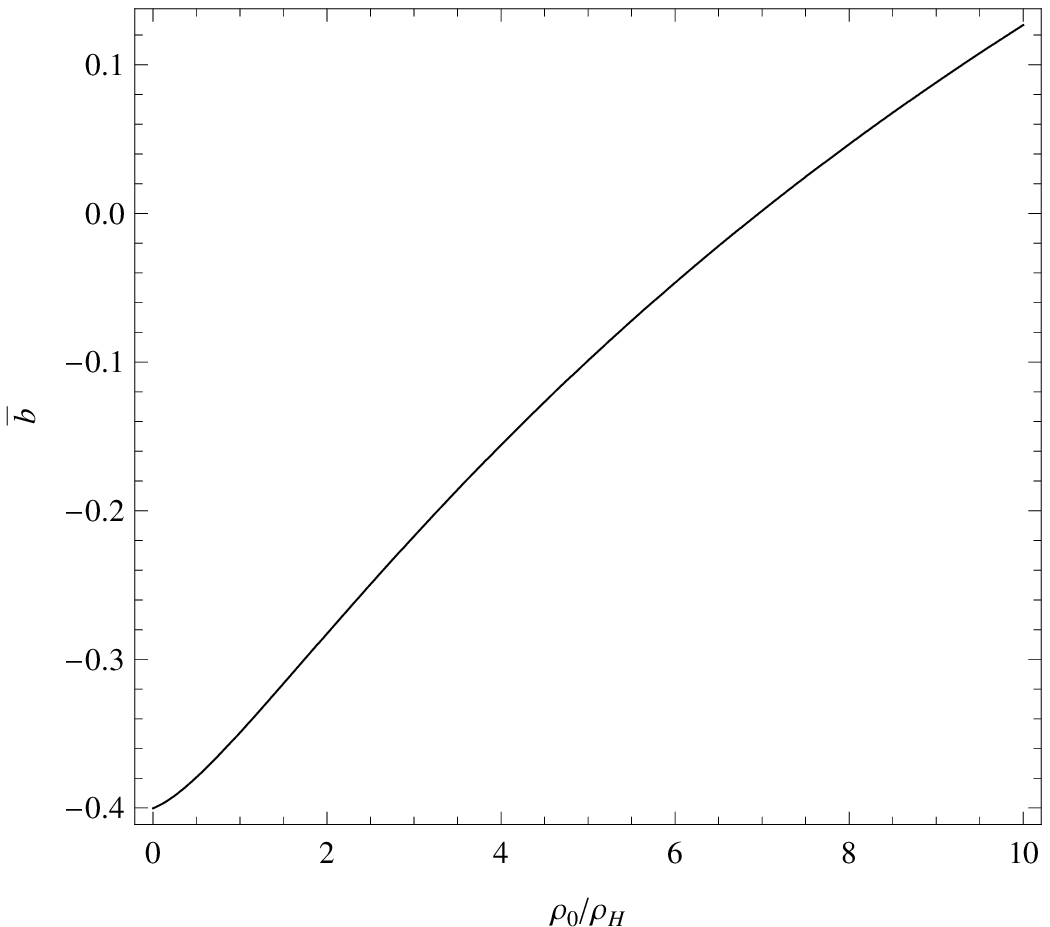}
\caption{Variation of the coefficients of the strong field limit
$\bar{a}$ (left) and $\bar{b}$ ( right) with parameter
$\rho_0/\rho_H$ in the squashed Kaluza-Klein black hole spacetime. }
\end{center}
\end{figure}
\begin{figure}[ht]
\begin{center}
\includegraphics[width=7cm]{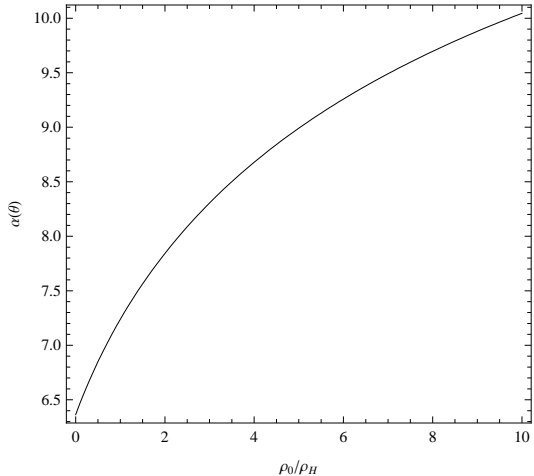}
\caption{Deflection angles in the squashed Kaluza-Klein black hole
spacetime evaluated at $u=u_{hs}+0.003$ as functions of
$\rho_0/\rho_H$. }
\end{center}
\end{figure}
Here $D_{OL}$ is the distance between observer and gravitational
lens. Making use of Eqs.(\ref{phs1}), (\ref{alf1}) and (\ref{coa1}),
we can study the properties of strong gravitational lensing in the
squashed Kaluza-Klein black hole spacetime. In the Figs. (2)-(4), we
plot the variations of the ratio $u_{hs}/\rho_H$, the coefficients
$\bar{a}$ and $\bar{b}$, and deflection angle $\alpha(\theta)$ with
the parameter of the extra dimension $\rho_0/\rho_H$. Obviously, as
$\rho_0$ tends to zero, these quantities reduce to those in the
four-dimensional Schwarzschild black hole spacetime. From Fig.(2),
we can see that with the increase of $\rho_0/\rho_H$, the ratio
$u_{hs}/\rho_H$ increases, which is different from that originated
from the charge in the Reissner-Norstr\"{om} black hole spacetime.
Moreover, from Fig.(3), we can find that the coefficients $\bar{a}$
and $\bar{b}$ increase with $\rho_0/\rho_H$. We also show the
deflection angle $\alpha(\theta)$ evaluated at $u=u_{hs}+0.003$ in
Fig.(4). It indicates that the presence of $\rho_0$ increases the
deflection angle $\alpha(\theta)$ for the light propagated in the
squashed Kaluza-Klein black hole spacetime. Comparing with those in
the four-dimensional Schwarzschild black hole spacetime, we could
extract information about the size of the extra dimension by using
strong field gravitational lensing.

\section{Observational gravitational lensing parameters}

Let us now study the effect of the scale parameter $\rho_0$ on the
observational gravitational lensing parameters. We start by assuming
that the gravitational field of the supermassive black hole at the
Galactic center of Milky Way can be described by the squashed
Kaluza-Klein black hole spacetime, and then  estimate the numerical
values for the coefficients and observables of gravitational lensing
in the strong field limit.

When the source, lens, and observer are highly aligned, the lens
equation in strong gravitational lensing can be approximated as
\cite{Bozza1}
\begin{eqnarray}
\beta=\theta-\frac{D_{LS}}{D_{OS}}\Delta\alpha_{n},
\end{eqnarray}
where $D_{LS}$ is the distance between the lens and the source,
$D_{OS}=D_{LS}+D_{OL}$, $\beta$ is the angular separation between
the source and the lens, $\theta$ is the angular separation between
the imagine and the lens, $\Delta\alpha_{n}=\alpha-2n\pi$ is the
offset of deflection angle, and $n$ is an integer. The $n$-th image
position $\theta_n$ and the $n$-th image magnification $\mu_n$ can
be approximated as
\begin{eqnarray}
\theta_n=\theta^0_n+\frac{u_{hs}(\beta-\theta^0_n)e^{\frac{\bar{b}-2n\pi}{\bar{a}}}D_{OS}}{\bar{a}D_{LS}D_{OL}},
\end{eqnarray}
\begin{eqnarray}
\mu_n=\frac{u^2_{hs}(1+e^{\frac{\bar{b}-2n\pi}{\bar{a}}})e^{\frac{\bar{b}-2n\pi}{\bar{a}}}D_{OS}}{\bar{a}\beta
D_{LS}D^2_{OL}},
\end{eqnarray}
respectively. The quantity $\theta^0_n$ is the image positions
corresponding to $\alpha=2n\pi$. In the limit $n\rightarrow \infty$,
the relation between the minimum impact parameter $u_{hs}$ and the
asymptotic position of a set of images $\theta_{\infty}$ can be
expressed as
\begin{eqnarray}
u_{hs}=D_{OL}\theta_{\infty}.\label{uhs1}
\end{eqnarray}
In order to obtain the coefficients $\bar{a}$ and $\bar{b}$, one
needs to separate at least the outermost image from all the others.
As in Refs.\cite{Bozza2,Bozza3},  we consider here the simplest case
in which only the outermost image $\theta_1$ is resolved as a single
image and all the remaining ones are packed together at
$\theta_{\infty}$. Thus the angular separation between the first
image and other ones can be expressed as
\begin{eqnarray}
s=\theta_1-\theta_{\infty},
\end{eqnarray}
and the ratio of the flux from the first image and those from the
all other images is given by
\begin{eqnarray}
\mathcal{R}=\frac{\mu_1}{\sum^{\infty}_{n=2}\mu_n}.
\end{eqnarray}
For highly aligned source, lens, and observer geometry, these
observables can be simplified as \cite{Bozza2,Bozza3}
\begin{eqnarray}
&s&=\theta_{\infty}e^{\frac{\bar{b}-2\pi}{\bar{a}}},\nonumber\\
&\mathcal{R}&= e^{\frac{2\pi}{\bar{a}}}.\label{ss1}
\end{eqnarray}
Thus, through measuring $s$, $\mathcal{R}$, and $\theta_{\infty}$,
we can obtain the strong deflection limit coefficients $\bar{a}$,
$\bar{b}$ and the minimum impact parameter $u_{hs}$. Comparing their
values with those predicted by the theoretical models, we could
detect the size of the extra dimension.
\begin{figure}[ht]
\begin{center}
\includegraphics[width=5cm]{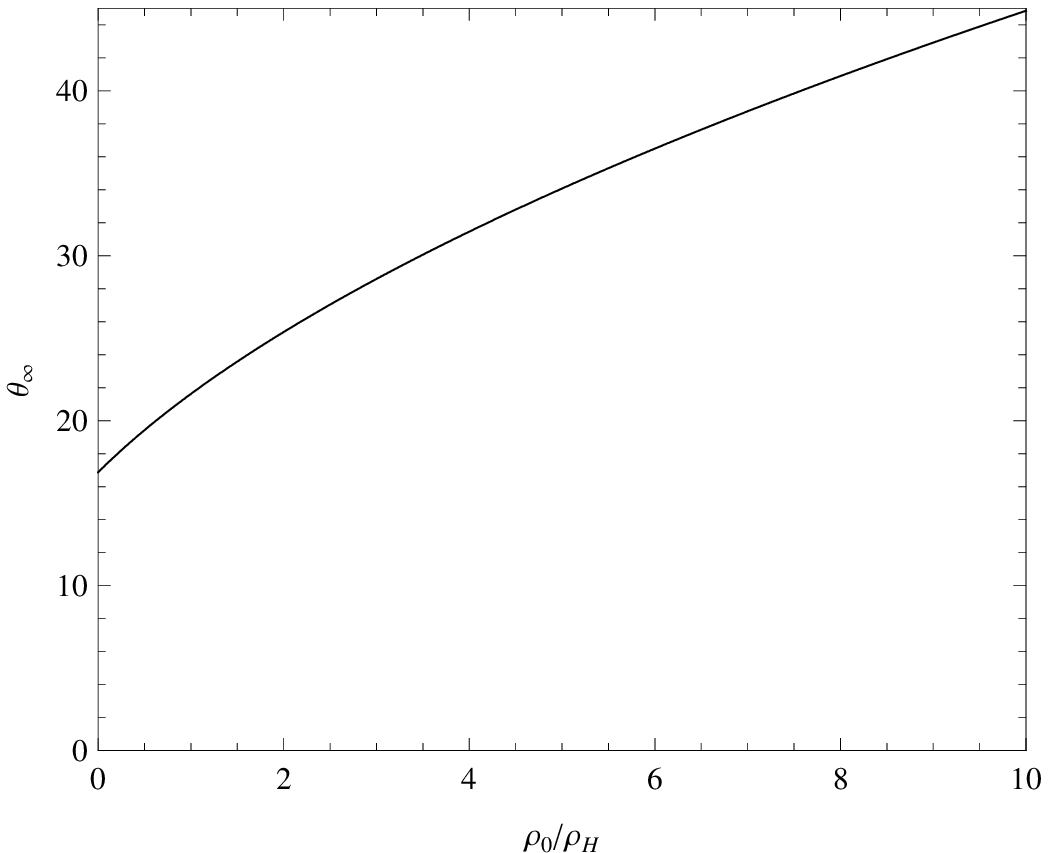}\;\;\;\;\;
\includegraphics[width=5cm]{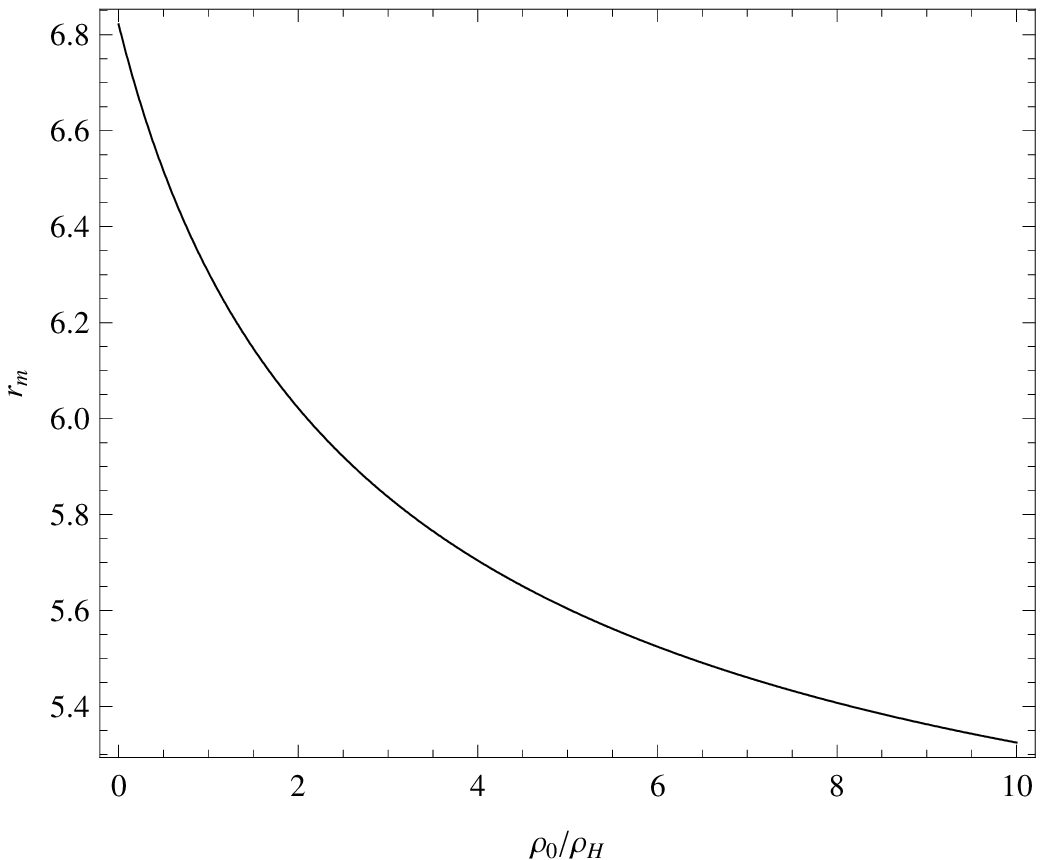}\;\;\;\;\;
\includegraphics[width=5cm]{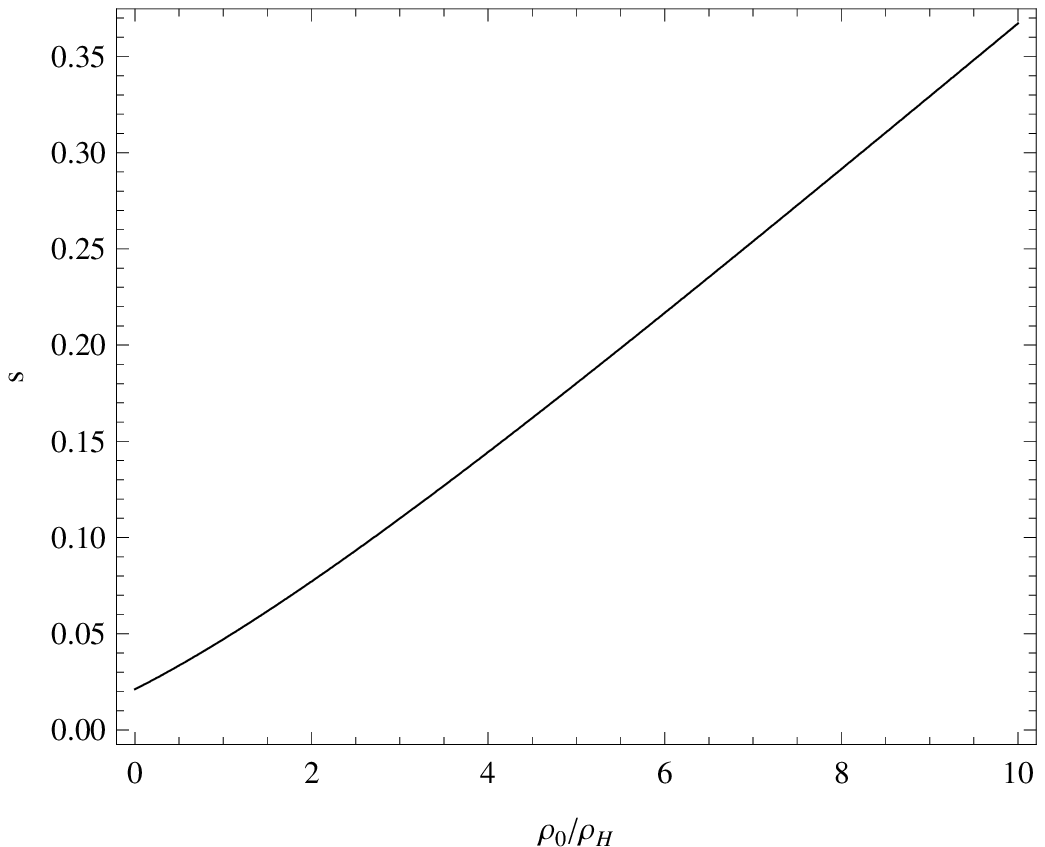}
\caption{Gravitational lensing by the Galactic center black hole.
Variation of the values of the angular position $\theta_{\infty}$
and the relative magnitudes $r_m$ with parameter $\rho_0/\rho_H$ in
the squashed Kaluza-Klein black hole spacetime.}
\end{center}
\label{f2}
\end{figure}
With the help of Eqs. (\ref{coa1}), (\ref{uhs1}) and in combination
with $\rho_H=2G_4M$, we can find that
\begin{eqnarray}
\theta_{\infty}=\frac{u_{hs}}{D_{OL}}=\frac{\sqrt{2}}{2}
\bigg[27+18\frac{\rho_0}{\rho_H}-\frac{\rho^2_0}{\rho^2_H}+\sqrt{\bigg(\frac{\rho_0}{\rho_H}+1\bigg)}
\bigg(\frac{\rho_0}{\rho_H}+9\bigg)^{3/2}\bigg]^{1/2}\frac{G_4M}{D_{OL}}.
\label{st}
\end{eqnarray}
The mass of the central object of our Galaxy is estimated to be
$2.8\times 10^6M_{\odot}$ and its distance is around $8.5kpc$, so
that the ratio of the mass to the distance $G_4M/D_{OL}
\approx1.574\times10^{-11}$ \cite{Vir1}. Here $D_{OL}$ is the
distance between the lens and the observer in the $\rho$
coordination rather than that in $r$ coordination because that in
the five-dimensional spacetime the dimension of the black hole mass
$M$ is the square of that in the polar coordination $r$. Making use
of Eqs. (\ref{st}), (\ref{coa1}) and  (\ref{ss1})  we can estimate
the values of the coefficients and observables for gravitational
lensing in the strong field limit. For the different
$\rho_0/\rho_H$, the numerical value for the relative minimum impact
parameter $u_{hs}/\rho_H$, the angular position of the relativistic
images $\theta_{\infty}$, the angular separation $s$ and the
relative magnification of the outermost relativistic image with the
other relativistic images $r_{m}$ (which is related to $\mathcal{R}$
by $r_m=2.5\log{\mathcal{R}}$) are listed in the Table I. The
dependence of these observables on the parameter $\rho_0/\rho_H$ are
also shown in Fig. (5).
\begin{table}[h]
\begin{center}
\begin{tabular}{|c|c|c|c|c|c|c|}
\hline \hline &&&&&& \\ $\rho_0/\rho_H$ &$\theta_{\infty}
$($\mu$arcsec)&\; $s$ ($\mu$arcsec) \;\; & $r_m$(magnitudes)
&\;\;\;\;$u_{hs}/\rho_H$\;\;\;\; &
\;\;\;\;\;\;\;\;$\bar{a}$\;\;\;\;\;\;\;\; &\;\;\;\;\;\;\;\;
$\bar{b}$\;\;\;\;\;\;\;\; \\
\hline
 0& 16.870& 0.02112& 6.8219&2.598& 1.000& -0.4002 \\
 \hline
0.1&17.420&0.02346&6.7497&2.683& 1.011&-0.3974\\
\hline
0.2& 17.949&0.02588&6.6838&2.764&1.021&-0.3938 \\
\hline
0.3& 18.458&0.02835&6.6234&2.843&1.030&-0.3895\\
 \hline
0.4 &18.950&0.03088&6.5678&2.918&1.039&-0.3847\\
 \hline
0.5&19.427&0.03346&6.5162&2.992&1.047&-0.3794
 \\
\hline\hline
\end{tabular}
\end{center}
\label{tab1} \caption{Numerical estimation for main observables and
the strong field limit coefficients for the black hole at the center
of our Galaxy, which is supposed to be described by the squashed
Kaluza-Klein black hole spacetime. $\rho_0$ is the parameter of the
extra dimension, $r_m=2.5\log{\mathcal{R}}$.}
\end{table}
Obviously, our results reduce to those in the four-dimensional
Schwarzschild black hole spacetime as $\rho_0=0$. From Table I and
Fig. (5), we find that with the increase of $\rho_0$, the relative
minimum impact parameter $u_{hs}/\rho_H$, the angular position of
the relativistic images $\theta_{\infty}$ and  the angular
separation $s$ increase, while the relative magnitudes $r_m$
decrease. This information could help us to detect the extra
dimension in the future.

\section{summary}

Gravitational lensing in the strong field limit provides a
potentially powerful tool to identify the nature of black holes in
the different gravity theories. The extra dimension is one of the
important predictions in string theory, which is believed to be a
promising candidate for the unified theory of everything. In this
paper we have investigated strong field lensing in the squashed
Kaluza-Klein black hole spacetime and found that the size of the
extra dimension imprints in the radius of the photon sphere, the
deflection angle, the angular position and magnification of the
relativistic images. The model was applied to the supermassive black
hole in the Galactic center. Our results show that with the increase
of the parameter $\rho_0/\rho_H$ the relative minimum impact
parameter $u_{hs}/\rho_H$, the angular position of the relativistic
images $\theta_{\infty}$ and the angular separation $s$ increase,
while the relative magnitudes $r_m$ decrease. This may offer a way
to detect the extra dimension by astronomical instruments in the
future.

\begin{acknowledgments}
This work was  partially supported by the National Natural Science
Foundation of China under Grant No.10875041,  the Program for
Changjiang Scholars and Innovative Research Team in University
(PCSIRT, No. IRT0964) and the construct program of key disciplines
in Hunan Province. J. Jing's work was partially supported by the
National Natural Science Foundation of China under Grant
No.10675045, No.10875040 and No.10935013; 973 Program Grant No.
2010CB833004 and the Hunan Provincial Natural Science Foundation of
China under Grant No.08JJ3010.
\end{acknowledgments}

\end{document}